\documentclass[journal,twoside,web]{ieeecolor}
\usepackage{jsen}
\usepackage{amsmath,amssymb,booktabs,graphicx,cite,siunitx,balance,placeins,needspace,wrapfig,textcomp}
\allowdisplaybreaks[1]
\graphicspath{{figures/}}
\title{Side-Vent Design for Microphone-Based Mechanomyography Sensors: Modeling and Experimental Evaluation of Contact Dependence}
\author{Cong Liu, Xiao Jin, and Zhenhua Yu
\thanks{This work has been submitted to the IEEE for possible publication. Copyright may be transferred without notice, after which this version may no longer be accessible. Manuscript received XX XXXX 2026. This work was supported in part by Capital Equipment funding from the University of Aberdeen. (Corresponding author: Zhenhua Yu.)}
\thanks{This work involved human subjects in its research. Approval of all ethical and experimental procedures and protocols was granted by the University of Aberdeen Ethics Committee (approval ID: 8891100), and informed consent was obtained.}
\thanks{Cong Liu is with the Department of Computer Science, University of Aberdeen, Aberdeen AB24 3UE, U.K.}
\thanks{Xiao Jin is with the Dyson School of Design Engineering, Imperial College London, London SW7 2DB, U.K.}
\thanks{Zhenhua Yu is with the Department of Computer Science, University of Aberdeen, Aberdeen AB24 3UE, U.K., and also with the Dyson School of Design Engineering, Imperial College London, London SW7 2DB, U.K. (e-mail: zhenhua.yu@abdn.ac.uk).}}

\definecolor{abstractbg}{rgb}{0.89804,0.94510,0.83137}
\begin{document}
\IEEEtitleabstractindextext{%
\fcolorbox{abstractbg}{abstractbg}{%
\begin{minipage}{\textwidth}%
\begin{wrapfigure}[12]{r}{3in}%
\includegraphics[width=3in]{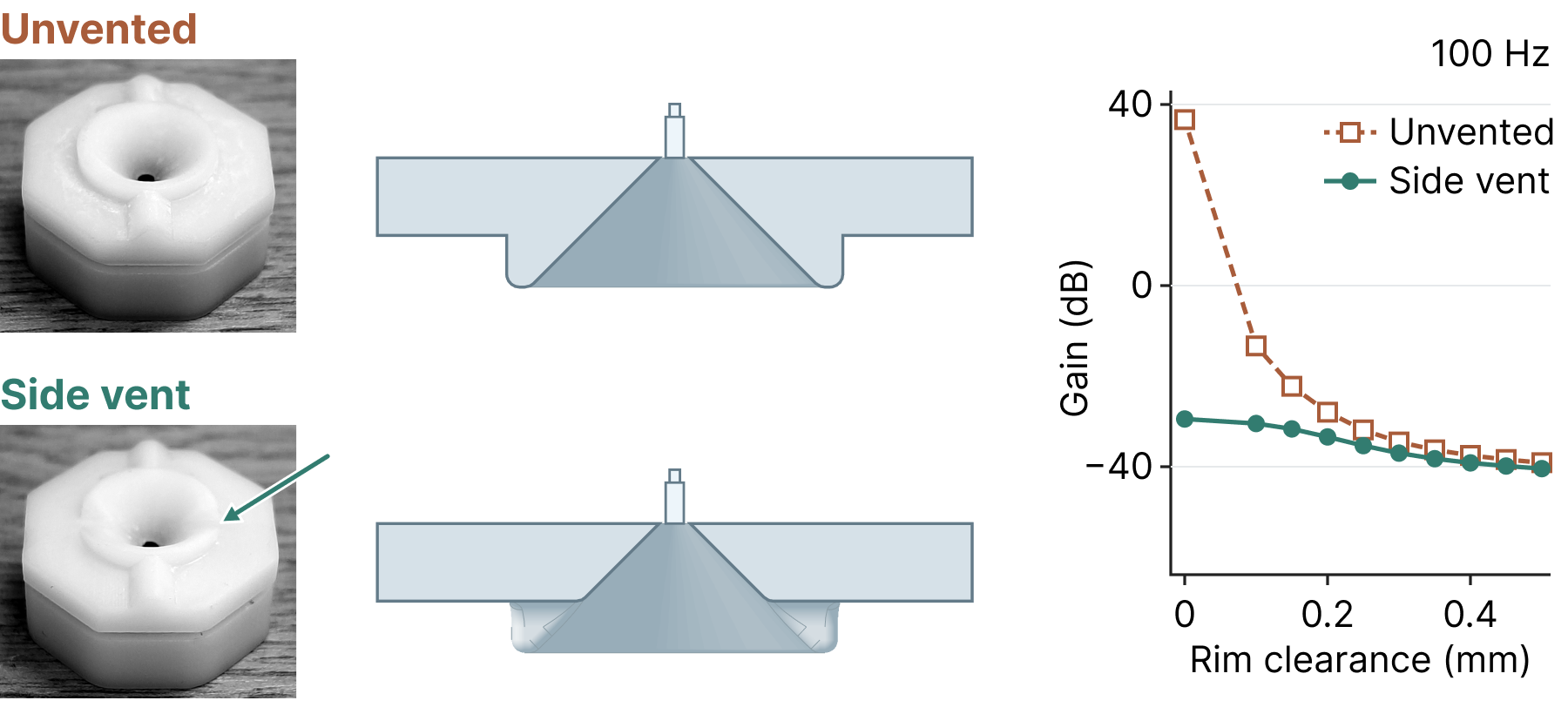}%
\end{wrapfigure}%
\begin{abstract}
Wearable muscle sensing for rehabilitation, assistive robotics, and electrical-stimulation feedback needs mechanomyography sensors less sensitive to fit changes during wear. In microphone-based mechanomyography, cavity pressure depends on rim contact, a dependence housing design has rarely targeted. This study evaluates a side-vented housing, whose vent opens a pressure-relief path in parallel with the rim to lower contact dependence, and quantifies the acoustic-gain cost. Thermoviscous finite-element simulations compared twenty cavity configurations under prescribed source motion. Paired bench sweeps over 10--200~Hz on both housings compared direct contact with a Gap-ring spacer, a radially grooved spacer ring. At 100~Hz, side venting reduced the simulated acoustic-gain variation across 0.10--0.50~mm rim clearances from 25.8 to 10.0~dB and cost 5.5~dB of gain at 0.20~mm clearance. Velocity fields and branch-flow calculations identified this path. At the higher speaker drive level, the Gap-ring spacer caused median attenuations of 13.6~dB (vented) and 30.1~dB (unvented) relative to direct contact. In this front end, a 35.9-dB signal-gain increase at 100~Hz raised the calculated operational-amplifier voltage-noise gain by 1.6~dB. A single-participant demonstration of paced elbow flexion and extension compared envelope timing and spectra between housings, finding similar envelope timing (correlation 0.935) and broadly similar spectra (correlation 0.956). Vent area trades acoustic gain against contact dependence and can therefore serve as a housing design parameter.
\end{abstract}

\end{minipage}}}

\maketitle

\section{Introduction}
\IEEEPARstart{W}{earable} mechanomyography (MMG) supports exoskeleton assistance \cite{jin2026exoskeleton}, human--machine and human--robot interaction \cite{castillo2021mmgforce,sheng2021integrated,jin2026multisensor}, prosthetic gesture recognition \cite{moqadam2022hybrid}, and muscle activity assessment \cite{kimoto2023wireless,islam2013mechanomyogram}. MMG records the low-frequency surface oscillations of contracting muscle fibers \cite{frangioni1987mechanism,orizio1993muscle}. In microphone-based MMG, skin-surface motion drives pressure fluctuations in an acoustic cavity coupled to the microphone diaphragm \cite{jin2026multisensor,paszkiewicz2022microphone}. With an open cavity, the housing rim rests on the skin, and the signal reflects cavity geometry, rim contact, and the electronic readout. On the limb, rim contact changes during movement and long-term wear. A wearable MMG sensor must therefore produce sufficient output and keep its response consistent under these changes.

Prior microphone-based MMG sensors focused on the acoustic cavity \cite{islam2013sensor}. Courteville \emph{et al.} developed a high-sensitivity cavity sensor \cite{courteville1998mmg}, and Watakabe \emph{et al.} tested condenser microphones with different air chambers \cite{watakabe2001mmg}. Posatskiy and Chau compared cylindrical and conical chambers \cite{posatskiy2012chamber}, and Yacoub \emph{et al.} examined how resonance-chamber geometry and dimensions affect the signal \cite{yacoub2023chamber}. A Mylar membrane closes the skin-facing opening in the Posatskiy--Chau chamber and later armband and gait sensors \cite{posatskiy2012chamber,castillo2021mmgforce,paszkiewicz2022microphone}, and Jin \emph{et al.} filled their chamber with damping silicone \cite{jin2026multisensor}. The chamber studies selected geometries for acoustic gain and frequency-response flatness. Most did not examine how the housing response changes with rim contact. Membranes and silicone keep skin contact outside the cavity boundary, whereas rim contact acts directly on open cavities. In the present open-cavity housing, rim contact can alter the acoustic boundary and microphone pressure.

Other studies examined system-level contact effects. Mancero Castillo \emph{et al.} evaluated how normal force affects hand-gesture classification with an acoustic MMG armband \cite{castillo2021mmgforce}. Jin \emph{et al.} tested capacitive-microphone MMG arrays under sweat, increased strap pressure, and loose attachment \cite{jin2026multisensor,jin2026exoskeleton}. Signal quality held under sweat and strap pressure but degraded under loose attachment, and mechanical stability mattered more than ambient noise \cite{jin2026multisensor}. These studies did not isolate housing acoustics, and no housing modification has been tested to reduce contact dependence. Contact dependence is the sensitivity of acoustic gain to rim contact, quantified by the open-clearance span in simulation (Section~III) and paired Direct-to-Gap-ring attenuation on the bench (Section~V). The Gap-ring is a grooved spacer ring that lifts the rim to emulate loss of rim contact.

Vents, leakage paths and sealing set the low-frequency response of small acoustic cavities. In hearing-aid earmolds, vent dimensions set the low-frequency reduction, and earplug leakage limits attenuation \cite{stuart1992probe,kuk2009comparison,henriksen2008using}. Fit-dependent leakage changes the response of in-ear microphones for wearable sensing \cite{demirel2023unobtrusive}. Sealing the ear canal raises its acoustic impedance and low-frequency pressure through the occlusion effect \cite{carillo2020theoretical,denk2023occlusion}. In miniature capacitive microphones, electrical and acoustic time constants govern low-frequency sensitivity \cite{shin2020modeling}, and vent and leak paths between the diaphragm and back cavity set the low-frequency corner \cite{peng2025analysis}. In body-sound sensing, a through-hole in the stethoscope diaphragm changes how the air column couples to the skin \cite{gaudio2025diaphragm}. In an open-cavity MMG housing, the rim clearance is a leakage path, and a side vent would act like these vents. Such a parallel pressure-relief path may lower contact dependence, because rim-clearance changes would alter acoustic gain less. This relief also lowers microphone pressure and thus acoustic gain.

This study tests this idea with a side-vented housing and quantifies the acoustic-gain cost. The evaluation must address three questions. First, a change in rim contact alters the acoustic path and the mechanical input to the cavity, so a bench test alone cannot isolate the vent's acoustic effect. Simulation under prescribed source motion isolates it, while bench tests measure the assembled sensor. Second, electronic gain must offset the lost acoustic gain. Usability depends on two criteria: the front-end noise cost of this added gain, and whether the recorded signal remains detectable. Third, the housing change must preserve the movement-related MMG signal, meaning its timing and spectral shape during muscle activity.

The contributions are:
\begin{enumerate}
\setlength{\itemsep}{0pt}\setlength{\parsep}{0pt}\setlength{\topsep}{2pt}
\item A thermoviscous finite-element model under prescribed source motion isolates the acoustic effect of rim clearance across twenty cavity configurations, with branch-flow and vent-area analyses. Paired bench sweeps compare direct contact with a Gap-ring spacer at fixed front-end gain within each housing. The side vent's parallel pressure-relief path lowers contact dependence in simulation (smaller open-clearance span) and on the bench (lower paired attenuation), making vent area a design parameter trading acoustic gain against contact dependence.
\item Calculations for this front end at two gain settings show that a large signal-gain increase raises op-amp voltage-noise gain slightly. On the bench, the Gap-ring lowers drive-frequency prominence less in the vented than in the unvented housing. Front-end gain restores the ADC signal level the vent removes at a small calculated op-amp noise cost, and acoustic gain sets the signal-to-noise ratio relative to microphone self-noise.
\item A single-participant on-body comparison during paced elbow movement showed similar envelope timing and broadly similar spectra for both housings, supporting on-body use of the vented housing.
\end{enumerate}

\section{Sensor Architecture and Housing Variants}
\subsection{Integrated module}
\begin{figure}[!t]
\centering
\includegraphics[width=\columnwidth]{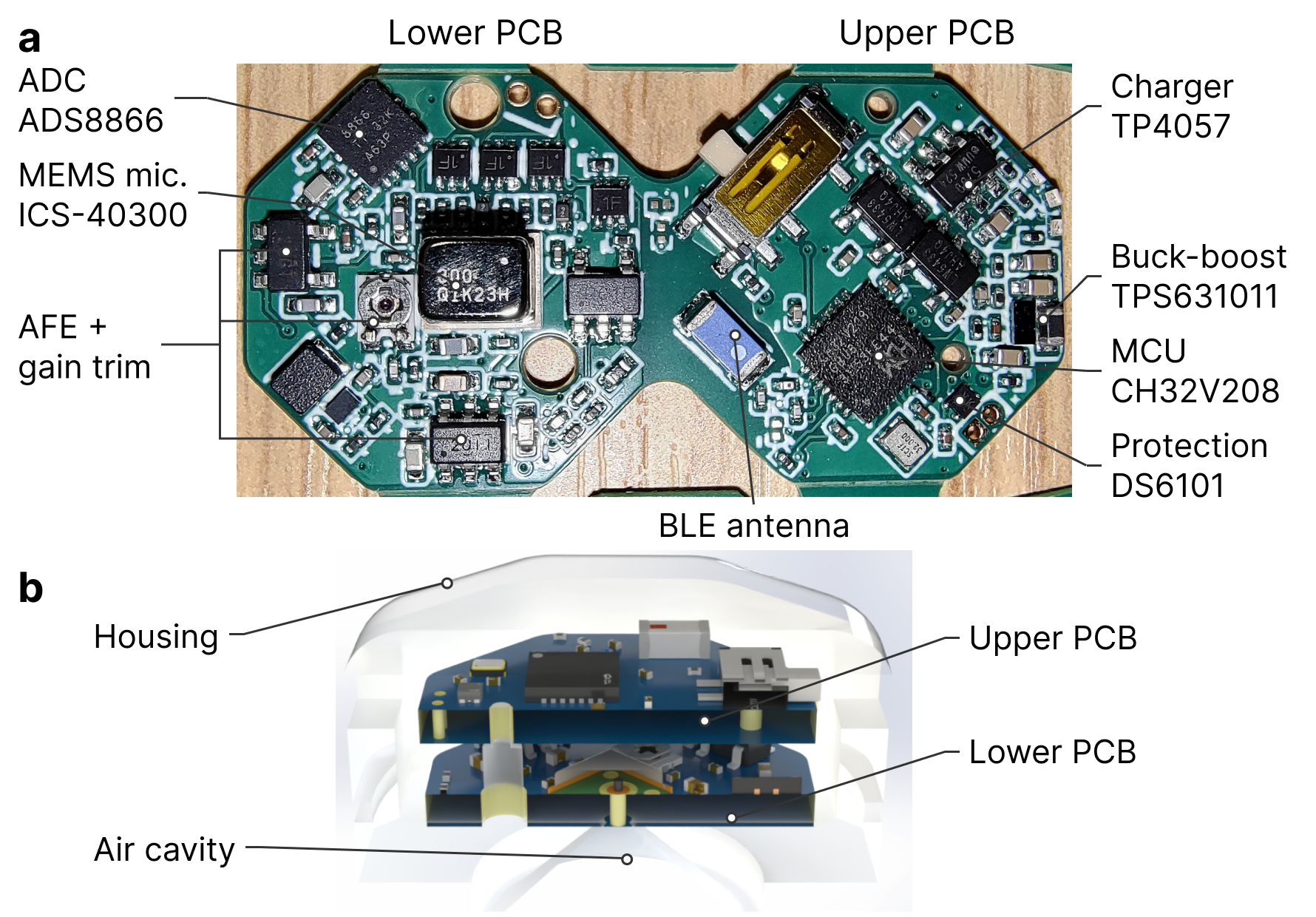}
\caption{Sensor hardware and assembly. (a) Lower and upper PCBs with the microphone, analog front end (AFE), gain adjustment, ADC, controller, antenna and power-management components labeled. (b) Cutaway view showing the housing, stacked PCBs and acoustic cavity. The two panels use different scales.}
\label{fig:system}
\end{figure}

The sensor consists of a body-facing acoustic housing and two stacked printed circuit boards (PCBs), as shown in Fig.~\ref{fig:system}. The lower PCB carries the ICS-40300 microelectromechanical systems (MEMS) microphone (TDK InvenSense) and analog front end, while the upper PCB carries the CH32V208 controller (WCH) and the power and communication circuits. The ADS8866 analog-to-digital converter (ADC, Texas Instruments) digitized the conditioned microphone signal in both the bench sweeps (Sections~V-A--V-C) and on-body recordings (Section~V-D). The CH32V208 transmitted the samples to the recording computer via Bluetooth Low Energy (BLE). Section~IV details the readout circuit and its calculated transfer characteristics.

\subsection{Housing variants and contact perturbation}
\begin{figure}[!t]
\centering
\includegraphics[width=\columnwidth]{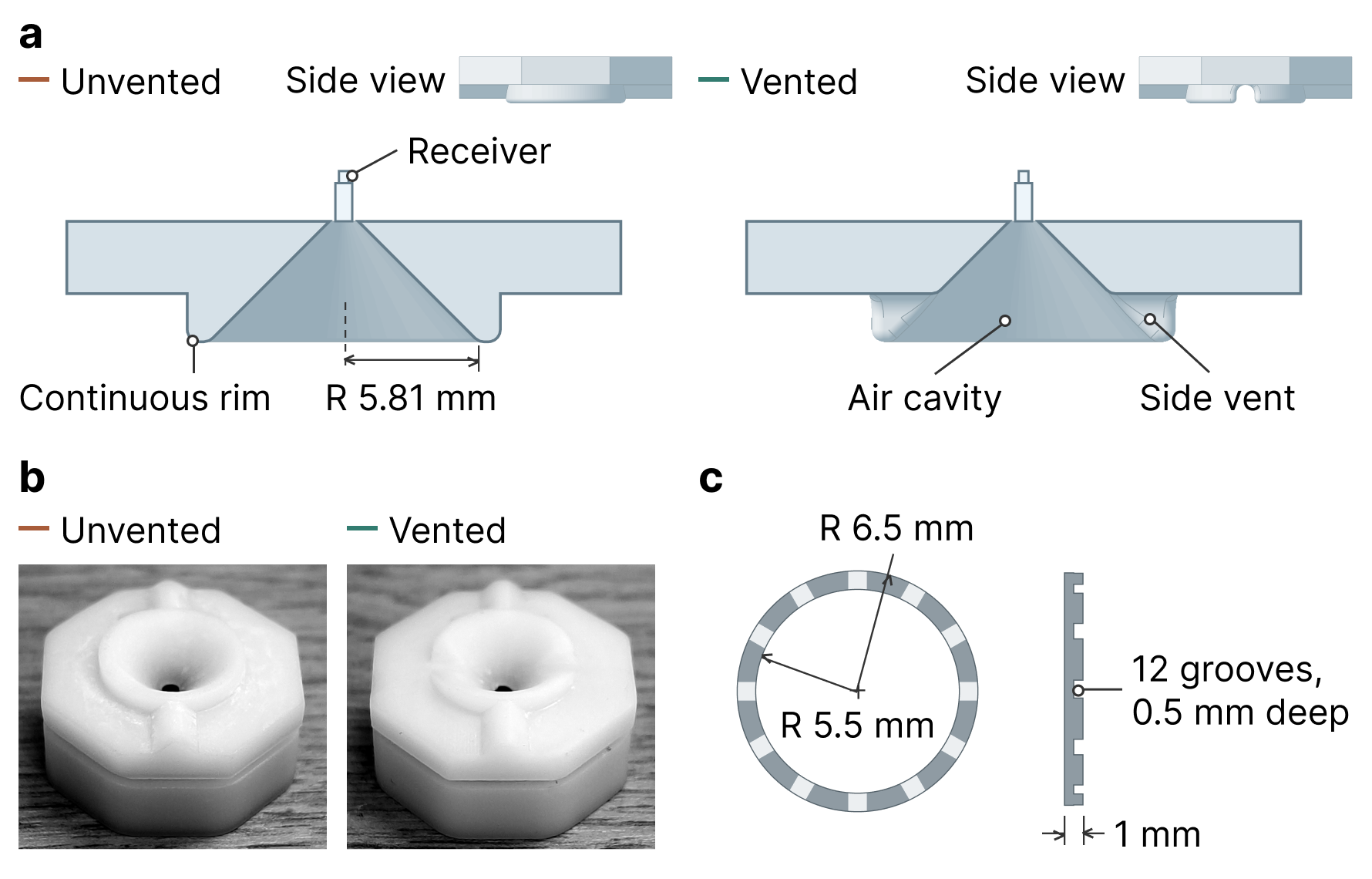}
\caption{Housing variants and contact spacer. (a) Cutaway and side views of the unvented and vented housings, with the receiver (microphone inlet), acoustic cavity, continuous rim and side vent labeled. (b) Photographs of the fabricated housings. (c) Plan and side views of the idealized Gap-ring spacer, showing its dimensions and radial grooves. Brown and green denote the unvented and vented housings, respectively.}
\label{fig:housing}
\end{figure}

Both housings share the same central cavity and acoustic path to the microphone. The vented housing adds two diametrically opposite side vents to the exterior (Fig.~\ref{fig:housing}(a)), and the unvented housing has a continuous rim. The computer-aided design (CAD) model specifies a nominal cavity height of 5.0~mm and a bottom inner radius of 5.81~mm after filleting. A 0.70-mm-diameter bore through the PCB connects the cavity to the 0.40-mm-diameter microphone inlet. Fig.~\ref{fig:housing}(b) shows the fabricated housings.

The rim contacted the contact surface (Section~V-A) directly (Direct) or through a removable spacer (Gap-ring). The housings and spacer were 3D printed in resin. The idealized spacer in Fig.~\ref{fig:housing}(c) is 1~mm thick, with inner and outer radii of 5.5 and 6.5~mm and twelve radial grooves extending 0.5~mm into it. The Gap-ring lifts the rim off the contact surface, and its grooves open outward leakage channels, emulating loss of rim contact. A crossed design tested each housing under both contact conditions. Simulations model rim-contact loss as rim clearance $g$ (Section~III), and bench tests use the Gap-ring (Section~V).

\section{Acoustic Design Comparison and Leakage Mechanism}
\subsection{Model and comparison conditions}
We solved a quarter-symmetry frequency-domain model in COMSOL Multiphysics~6.0 \cite{comsol60}. At 100~Hz, the viscous penetration depth in air is 0.22~mm, comparable to the rim clearances studied, so the near field uses the thermoviscous formulation \cite{stinson1991propagation,kampinga2010performance,berggren2018acoustic}. A thermoviscous near-field domain of radius 8~mm couples to an exterior pressure-acoustic domain of radius 20~mm through the built-in acoustic--thermoviscous acoustic boundary coupling, with a spherical-wave radiation condition at the outer hemispherical boundary. The cavity, PCB bore and microphone inlet follow the CAD geometry. Free-tetrahedral meshes without boundary-layer elements have maximum element sizes of 0.5~mm in the near field and 2~mm in the exterior domain (Section~III-C reports mesh, boundary-layer and domain checks). Near-field pressure uses linear interpolation, and velocity and temperature use quadratic serendipity interpolation. The exterior domain uses quadratic pressure interpolation. The baseline and vented models contain 35,000--87,000 elements and $1.7\times10^{5}$--$5.5\times10^{5}$ degrees of freedom.

A flat plane represents the contact surface, forms the lower fluid boundary and extends under the housing rim to the domain edge. Its central disc, the vibrating skin region, provides the prescribed source motion, while the rest of the plane is rigid. Stationary walls use no-slip isothermal conditions in the thermoviscous domain and sound-hard conditions in the exterior domain. Clearance $g$ is set by lowering the plane below the fixed rim and rebuilding the fluid domain. The gap and vents open to the exterior domain. An impedance boundary condition with $Z_m=\rho_0c_0$ is applied at the 0.40-mm-diameter inlet face. Air properties at 293.15~K: $\rho_0=1.204$~kg/m$^3$, $c_0=343$~m/s, dynamic viscosity $1.81\times10^{-5}$~Pa\,s, thermal conductivity 0.0257~W/(m\,K), specific heat capacity 1005~J/(kg\,K), specific-heat ratio 1.4.

For each configuration, the acoustic pressure-to-displacement transfer function is defined as
\begin{equation}
H(f,g)=\frac{\overline{p}_{\mathrm{mic}}(f,g)}{\overline{u}_{\mathrm{s}}(f,g)},
\end{equation}
where $f$ is frequency and $g$ is rim clearance. The numerator is the complex pressure averaged over the microphone termination face. The denominator is the source displacement averaged over the source surface from the solved field. Complex values are averaged before $|H|$ is computed. The gain is expressed as $20\log_{10}[|H|/(1~\mathrm{Pa}/\si{\micro\meter})]$ in decibels. The prescribed source displacement follows $u_z(r)=3u_0[1-(r/r_s)^2]^2$, where $r$ is radial distance from the source center, and its area average is $u_0$. This shape is chosen to represent skin motion that peaks at the center and vanishes at the disc edge, where the skin meets the rim region. For the baseline geometry, the source radius is $r_s=5.6$~mm and the mean displacement amplitude is $u_0=1$~nm. The model is linear, so $H$ is independent of $u_0$, which only sets the field scale. The profile is applied as the corresponding normal velocity, with tangential velocity set to zero. All housings and clearances share this profile and mean amplitude. For lateral and uniform scaling, the source radius $r_s$ scales with the cavity opening.

For each housing, we compare the transfer magnitude at $g=0.10$, 0.20 and 0.50~mm, the three open clearances solved for all 20 configurations. At each frequency, the difference between the highest and lowest gains defines the open-clearance span,
\begin{equation}
S(f)=20\log_{10}\left(\frac{\max_{g\in\{0.10,0.20,0.50\}}|H(f,g)|}{\min_{g\in\{0.10,0.20,0.50\}}|H(f,g)|}\right).
\end{equation}
A smaller $S$ means the gain varies less across these clearances. The $g=0$ condition represents ideal rim closure and is evaluated separately.

We test sensitivity to $Z_m=\rho_0c_0$, the characteristic impedance of air, used as a common reference termination. At 100~Hz, open-clearance transfer magnitudes stay below 0.22~Pa/\si{\micro\meter} for both housings. The unvented housing reaches 68~Pa/\si{\micro\meter} at rim closure. A rigid-wall termination changes open-clearance gains by at most 0.07~dB and unvented closure gain by 0.58~dB. Termination choice matters most in the sealed unvented condition.

\subsection{Design-space comparison}
We compared 20 housing configurations in the CAD-based model at 100~Hz, a representative frequency: the unvented baseline, the side-vented housing and 18 baseline variants. Eleven varied cavity profile, volume, height, PCB bore and shell material (two per geometric factor, three shell-material parameter sets). The other seven comprised three reduced-height designs, three lateral scales and one uniform scale. All configurations were evaluated at rim clearances $g=0$, 0.10, 0.20 and 0.50~mm, and the baseline and side-vented housing also at 0.15--0.45~mm in 0.05-mm steps. Section~III-C examines both housings' frequency dependence over 10--200~Hz.

The baseline cavity forms a 45$^{\circ}$ conical section extending from the filleted bottom opening to a 1.2-mm-diameter throat at the PCB bore. Its air volume is 183.87~mm$^3$, with the side vents adding a further 13.77~mm$^3$ in the vented housing.

Profile variants displace the connecting wall by up to 0.45~mm. Volume variants scale the cavity radially to 65\,\% or 135\,\% of the baseline volume. Height variants retain 45$^{\circ}$ wall slopes. Reduced-height designs (2.5, 1.5 and 1.0~mm) keep the bottom opening, source radius and throat at the PCB bore but replace the conical wall with a smooth curved wall. PCB-bore variants change only the hole diameter. Lateral scaling shrinks the opening and driven radius at fixed height. Uniform scaling shrinks all dimensions but retains the fabricated bore and inlet. Material values $(E,\rho_s,\nu)$ are $(2500~\mathrm{MPa},1250~\mathrm{kg/m^3},0.35)$ for polylactic acid (PLA), $(2000~\mathrm{MPa},1150~\mathrm{kg/m^3},0.35)$ for resin and $(3~\mathrm{MPa},1100~\mathrm{kg/m^3},0.49)$ for silicone, with the elastic shell fixed at its upper face and a loss factor of 0.05.

\begin{figure}[!t]
\centering
\includegraphics[width=\columnwidth]{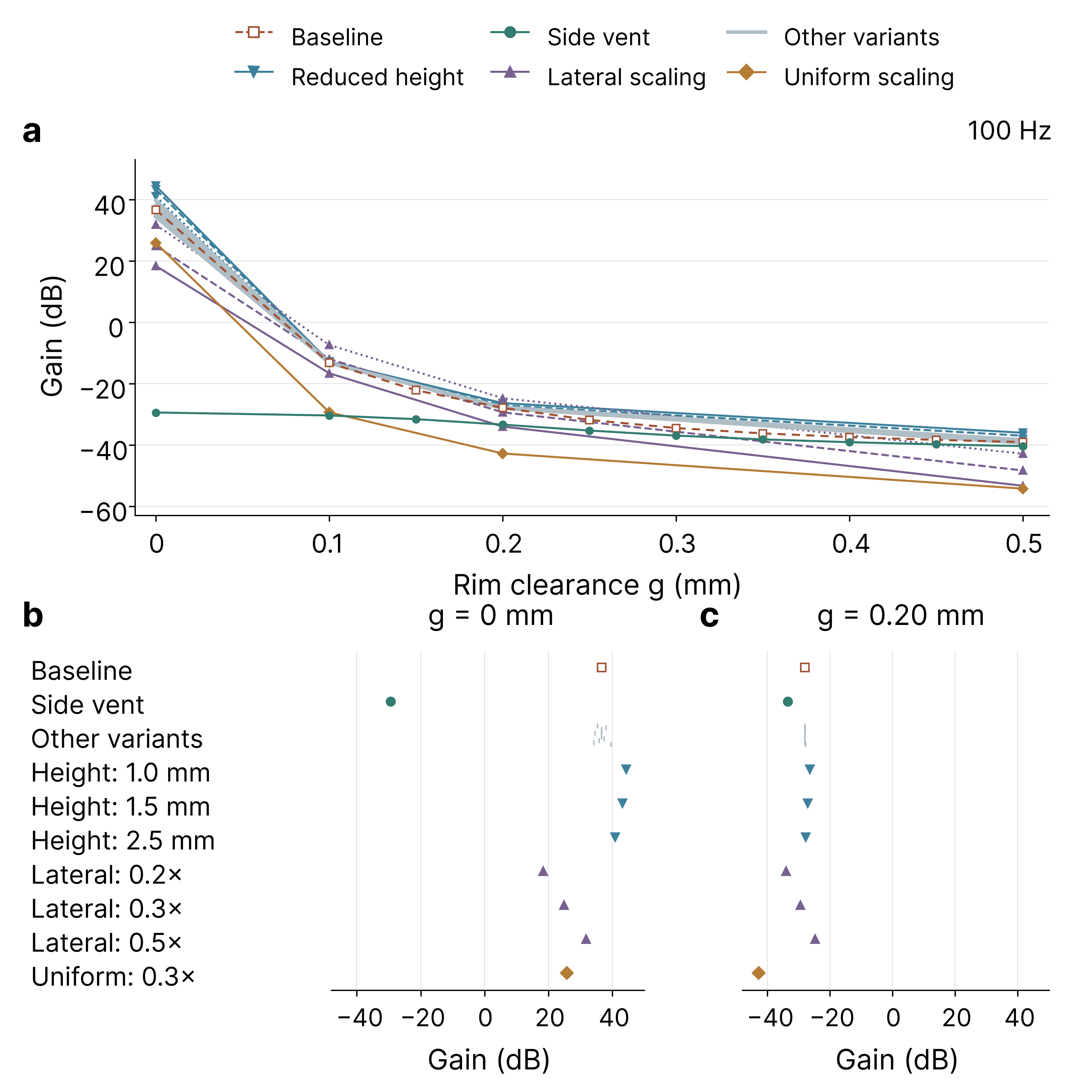}
\caption{Calculated acoustic gain for 20 cavity configurations at 100~Hz. (a) Gain versus rim clearance. (b,c) Gain at ideal rim closure and at $g=0.20$~mm, respectively, using matching gain scales and configuration order. Gain is expressed as $20\log_{10}[|H|/(1~\mathrm{Pa}/\si{\micro\meter})]$. Markers denote solved conditions, and lines connect them. The eleven modest geometry and material variants are each drawn as individual grey lines and grouped in the legend as Other variants. In the reduced-height group in (a), solid, dashed and dotted lines indicate heights of 1.0, 1.5 and 2.5~mm, respectively. In the lateral-scaling group, the same line styles indicate scale factors of 0.2, 0.3 and 0.5, respectively.}
\label{fig:gain}
\end{figure}

Table~\ref{tab:configs} summarizes the calculated gain and open-clearance span for the configurations. At 100~Hz, the side-vented housing has the lowest open-clearance span among all configurations (Fig.~\ref{fig:gain}). Its span is 9.98~dB, compared with 25.79~dB for the unvented baseline. This reduction comes with a gain loss of 5.45~dB at $g=0.20$~mm. The loss increases from 1.3~dB at $g=0.50$~mm to 17.1~dB at $g=0.10$~mm.

The eleven modest geometry and material variants differ from the baseline by up to about 3~dB at $g=0$ but at most 0.7~dB at open clearances. Reduced-height and scaled designs produce larger gain changes than these eleven variants, but their open-clearance spans remain above that of the side-vented housing. Reducing the height to 1.0~mm raises $G_0$ from 36.64 to 44.40~dB and the gain at $g=0.20$~mm by 1.55~dB. Lateral scaling by a factor of 0.5 raises gain at $g=0.20$~mm by 3.18~dB, yet across all three lateral scale factors the span increases to 35.47--36.68~dB. Uniform scaling by a factor of 0.3 lowers the gain at $g=0.20$~mm by 14.86~dB.

\begin{table}[!t]
\centering
\caption{Calculated Acoustic Gain and Open-Clearance Span for 20 Cavity Configurations at 100~Hz}
\label{tab:configs}
\footnotesize
\setlength{\tabcolsep}{3.2pt}
\begin{tabular}{@{}llrrrr@{}}
\toprule
Group & Setting & $G_{0}$ & $G_{0.20}$ & $\Delta G_{0.20}$ & $S$ \\
 & & (dB) & (dB) & (dB) & (dB) \\
\midrule
Reference & Unvented baseline & 36.64 & $-$27.98 & -- & 25.79 \\
 & Side vent & $-$29.47 & $-$33.43 & $-$5.45 & 9.98 \\
\addlinespace[1pt]
Profile & Inward & 37.39 & $-$27.95 & $+$0.03 & 25.65 \\
 & Outward & 35.87 & $-$28.01 & $-$0.03 & 25.91 \\
\addlinespace[1pt]
Volume & 65\,\% & 39.63 & $-$27.79 & $+$0.19 & 25.11 \\
 & 135\,\% & 34.30 & $-$28.06 & $-$0.08 & 26.10 \\
\addlinespace[1pt]
Height & 4\,mm & 38.10 & $-$27.94 & $+$0.04 & 25.60 \\
 & 6\,mm & 35.39 & $-$28.00 & $-$0.02 & 25.89 \\
\addlinespace[1pt]
PCB bore & 0.50\,mm & 36.63 & $-$28.01 & $-$0.03 & 25.79 \\
 & 0.90\,mm & 36.63 & $-$27.99 & $-$0.01 & 25.79 \\
\addlinespace[1pt]
Shell & PLA/resin & 36.64 & $-$27.98 & 0.00 & 25.79 \\
 & Silicone & 34.65 & $-$27.96 & $+$0.02 & 25.78 \\
\addlinespace[1pt]
Reduced height & 2.5\,mm & 40.96 & $-$27.69 & $+$0.29 & 24.88 \\
 & 1.5\,mm & 43.24 & $-$27.16 & $+$0.82 & 23.90 \\
 & 1.0\,mm & 44.40 & $-$26.43 & $+$1.55 & 23.20 \\
\addlinespace[1pt]
Lateral scale & 0.5 & 31.77 & $-$24.80 & $+$3.18 & 35.47 \\
 & 0.3 & 24.83 & $-$29.43 & $-$1.45 & 36.36 \\
 & 0.2 & 18.31 & $-$34.07 & $-$6.09 & 36.68 \\
\addlinespace[1pt]
Uniform scale & 0.3 & 25.83 & $-$42.84 & $-$14.86 & 24.88 \\
\bottomrule
\end{tabular}

\vspace{2pt}
\parbox{\columnwidth}{\footnotesize $G_0$ and $G_{0.20}$ denote acoustic gain at ideal rim closure and at $g=0.20$~mm, respectively. Gains are expressed as $20\log_{10}[|H|/(1~\mathrm{Pa}/\si{\micro\meter})]$. $\Delta G_{0.20}$ is the gain difference from the unvented baseline at $g=0.20$~mm. $S$ is the open-clearance span defined in (2), evaluated over $g=0.10$, 0.20 and 0.50~mm. The baseline has a cavity height of 5.0~mm, a PCB-bore diameter of 0.70~mm and a rigid shell. PLA and resin shells gave identical values to two decimals.}
\end{table}

\subsection{Flow mechanism and frequency dependence}
\begin{figure}[!t]
\centering
\includegraphics[width=\columnwidth]{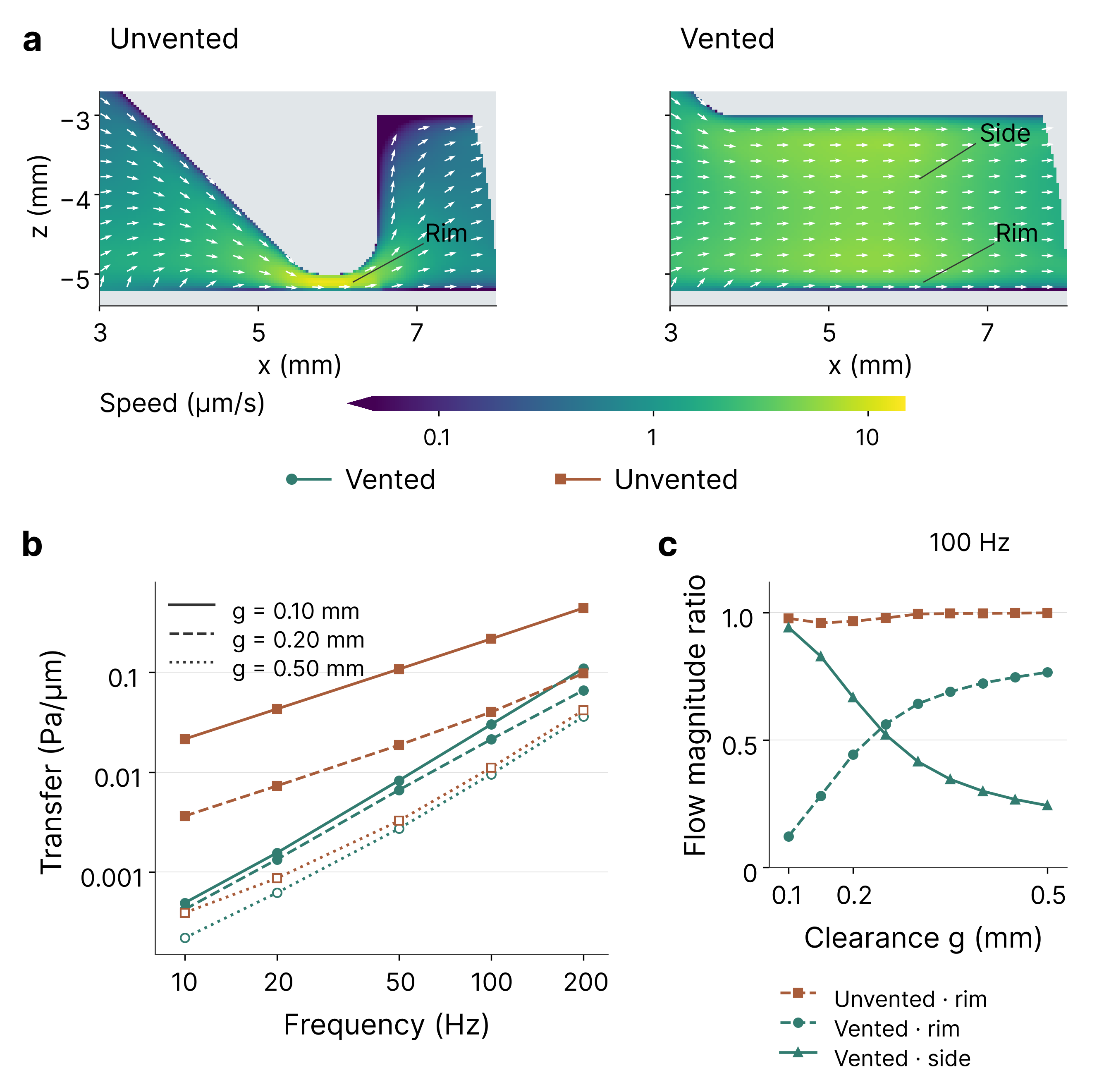}
\caption{Acoustic response and flow paths under prescribed source motion. (a) Computed velocity fields near the rim and side vent at 100~Hz and $g=0.20$~mm, shown on the $y=0.025$~mm plane. Both fields use the same spatial scale and logarithmic color scale. Arrows indicate the instantaneous in-plane velocity direction at a phase of $-90^{\circ}$. (b) Pressure-to-displacement transfer magnitude for the vented and unvented models at three open clearances and five frequencies. (c) Branch-flow magnitudes normalized by the input-flow magnitude at 100~Hz for nine clearances from 0.10 to 0.50~mm. The ratios use magnitudes of complex flows. Lines connect solved points.}
\label{fig:mechanism}
\end{figure}

At 100~Hz and $g=0.20$~mm, the velocity fields show an additional lateral flow branch through the side vent (Fig.~\ref{fig:mechanism}). To quantify this redistribution, we integrated the complex flow through a cylindrical cut at a radius of 6~mm and normalized each branch magnitude by the input-flow magnitude. For the vented housing, as the clearance increased from 0.10 to 0.50~mm, the side-vent ratio decreased from 0.942 to 0.243, while the rim ratio increased from 0.122 to 0.766. The branch phases differ, so these magnitude ratios do not sum to one. Finer clearance sampling placed the crossover between the two branch magnitudes at $g=0.20$--0.25~mm. Below this range, most flow leaves through the side vent. Above it, the rim branch is larger. At small clearances, the vent therefore sets the cavity pressure, so a change in rim clearance alters the gain less, which gives the vented housing its smaller open-clearance span.

We compared the unvented baseline and side-vented housing at 10, 20, 50, 100 and 200~Hz. As frequency increased from 10 to 200~Hz, the unvented open-clearance span decreased from 34.80 to 20.47~dB. The vented span remained between 7.02 and 9.98~dB across the five frequencies. The gain reduction associated with the side vent depended on both frequency and clearance. At $g=0.20$~mm, the vented gain was 18.68~dB below the baseline at 10~Hz and 3.38~dB below it at 200~Hz.

Mesh, boundary-layer and domain checks changed the gain by at most 0.371~dB. Changing the branch-flow integration grid changed the flow ratios by at most 0.135\% of input flow. Both are far smaller than the 15.8-dB difference in open-clearance span between the vented and unvented housings at 100~Hz.

\subsection{Effect of vent opening area}
The fabricated side vent is a rim notch, 2~mm wide and 2~mm high, with an arched roof and 0.5-mm mouth fillets. Its minimum cross-sectional area is 3.57~mm$^2$ per vent. This geometry defines the side-vented model in Table~\ref{tab:configs} and Fig.~\ref{fig:gain}. Because the notch spans the full lip height, we varied its width while keeping the height fixed at 2~mm. For this area sweep, the opening was represented by a straight prismatic slot without fillets. The five areas were 0.89, 1.79, 3.57, 7.14 and 14.28~mm$^2$ per vent, corresponding to 0.25--4 times the fabricated area.

Increasing the opening area reduced the open-clearance span $S$ from 17.51 to 4.08~dB, with reductions of 3.2--3.7~dB per doubling. Doubling the fabricated opening area from 3.57 to 7.14~mm$^2$ reduced $S$ by 3.2~dB and added 2.6~dB of gain loss. Doubling it again to 14.28~mm$^2$ produced a similar reduction in $S$ of 3.3~dB, but added 3.9~dB of gain loss. Each doubling thus gave a similar reduction in $S$ at a growing gain cost.

\begin{figure}[!t]
\centering
\includegraphics[width=\columnwidth]{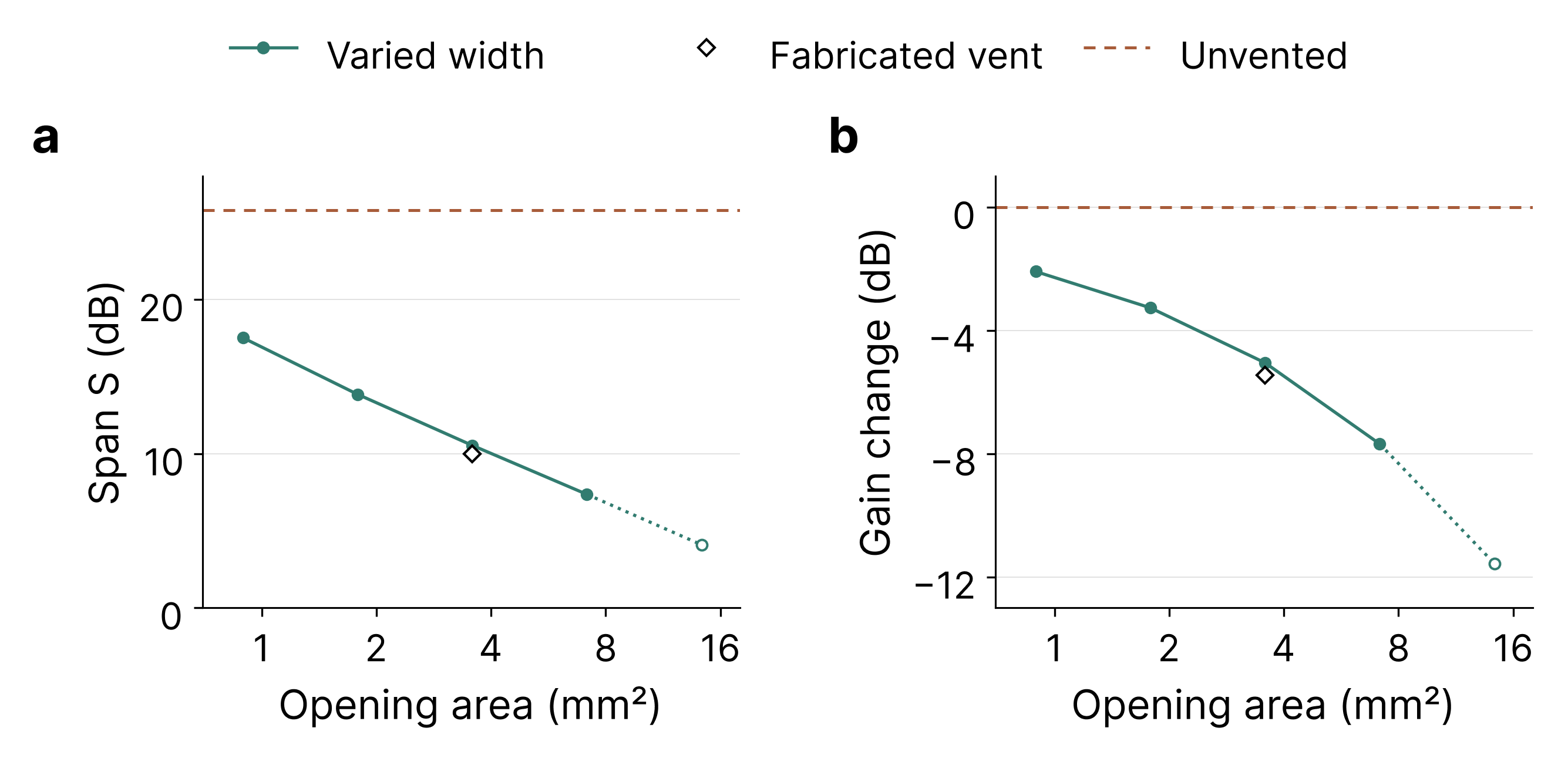}
\caption{Effect of vent opening area on open-clearance span and acoustic gain at 100~Hz. (a) Open-clearance span $S$ versus area per opening, with a base-2 logarithmic area axis. (b) Gain change $\Delta G_{0.20}$, the gain difference relative to the unvented housing at $g=0.20$~mm, versus area per opening. Filled markers and the green line represent the width sweep at a fixed height of 2~mm. The open diamond denotes the fabricated vent, and the dashed brown line denotes the unvented housing. The largest opening removes approximately $\pm40^{\circ}$ of rim on each side and has reduced local mesh quality. Its result is shown as indicative using an open marker and a dotted segment.}
\label{fig:ventarea}
\end{figure}

At the fabricated area, the prismatic-slot model and the filleted CAD vent differed by 0.55 and 0.40~dB in $S$ and $G_{0.20}$. At equal area, joint height-width scaling changed $S$ and $G_{0.20}$ by at most 0.7 and 0.05~dB relative to width-only scaling. Vent area dominated over aspect ratio in the cases tested. The largest vent removes about $\pm40^{\circ}$ of rim with reduced local mesh quality, as marked in Fig.~\ref{fig:ventarea}.

\section{Analog Readout and Calculated Transfer}
The reduced contact dependence of the side-vented housing comes at the cost of lower acoustic gain, motivating the use of greater electronic amplification. We calculated signal gain and op-amp noise at two gain settings to examine the readout implications of this tradeoff.
\subsection{Signal path}
\begin{figure}[!t]
\centering
\includegraphics[width=\columnwidth]{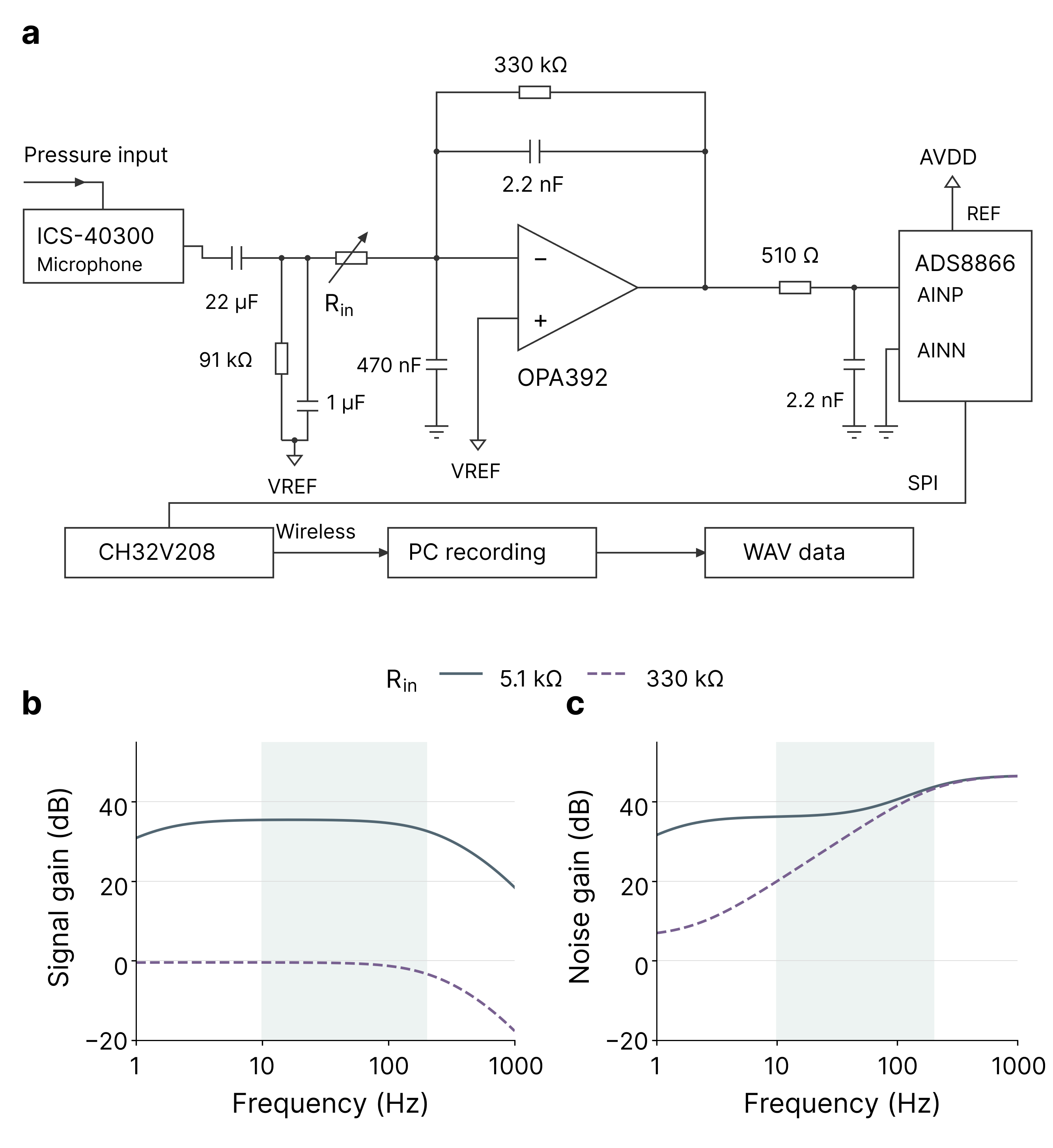}
\caption{Readout circuit and calculated gains. (a) Signal path comprising the microphone, adjustable amplifier, ADC and CH32V208 for wireless data transmission. VREF provides the analog bias, and AVDD serves as the ADC reference. (b) Signal gain from the microphone output to the ADC input. (c) Op-amp voltage-noise gain. Curves correspond to the two indicated input resistances $R_{\mathrm{in}}$. Shading marks the 10--200~Hz band. Calculations use a 200-$\Omega$ microphone source impedance and assume an ideal op-amp and an AC-stiff VREF.}
\label{fig:afe}
\end{figure}

Fig.~\ref{fig:afe} shows the signal path from the microphone to wireless transmission. The ICS-40300 output is AC-coupled through 22~\si{\micro\farad} and biased to VREF through 91~\si{k\ohm} and 1~\si{\micro\farad}. The input resistance $R_{\mathrm{in}}$ sets the gain of the OPA392 (Texas Instruments) inverting stage. The feedback network consists of a 330-k$\Omega$ resistor in parallel with a 2.2-nF capacitor. A 470-nF capacitor connects the summing node to ground and is included in the noise-gain calculation. The ADC input uses a 510-$\Omega$/\mbox{2.2-nF} network. The 16-bit ADS8866 has its AINN pin grounded, and its reference is connected to AVDD. The CH32V208 receives the digitized samples through a serial peripheral interface (SPI) and transmits them wirelessly.

\subsection{Calculated signal and noise gain}
The small-signal model assumes an ideal op-amp, an AC-stiff VREF and a microphone source impedance of $R_s=200~\Omega$. With $s=j2\pi f$, we define
\begin{align}
Z_f&=\frac{330~\mathrm{k\Omega}}{1+s(330~\mathrm{k\Omega})(2.2~\mathrm{nF})},\nonumber\\*
Z_L&=\left[\frac{1}{91~\mathrm{k\Omega}}+\frac{1}{R_{\mathrm{in}}}+s(1~\si{\micro\farad})\right]^{-1},\nonumber\\*
P&=\left[1+s(510~\Omega)(2.2~\mathrm{nF})\right]^{-1}.
\end{align}
The signal transfer from the microphone output to the ADC input is
\begin{equation}
G_s=-\frac{Z_f}{R_{\mathrm{in}}}\frac{Z_L}{R_s+1/[s(22~\si{\micro\farad})]+Z_L}P.
\end{equation}
For the op-amp voltage-noise calculation, we define $Y_g=[R_s+1/(s\,22~\si{\micro\farad})]^{-1}+(91~\mathrm{k\Omega})^{-1}+s(1~\si{\micro\farad})$ and $Y_i=1/R_{\mathrm{in}}$. The voltage-noise gain at the ADC input is
\begin{equation}
G_n=\left[1+Z_f\left(s(470~\mathrm{nF})+\frac{Y_iY_g}{Y_i+Y_g}\right)\right]P.
\end{equation}
These expressions account for input loading and the 470-nF capacitor at the summing node. The two input resistances, $R_{\mathrm{in}}=5.1~\mathrm{k\Omega}$ and $330~\mathrm{k\Omega}$, match the settings used for the on-body recordings in Section~V-D. The vented housing uses the higher-gain 5.1-k$\Omega$ setting. The unvented housing uses 330~k$\Omega$.

At 100~Hz, the calculated signal gain is 34.63~dB for $R_{\mathrm{in}}=5.1~\mathrm{k\Omega}$ and $-1.29$~dB for $R_{\mathrm{in}}=330~\mathrm{k\Omega}$. The corresponding op-amp voltage-noise gains are 40.59 and 39.00~dB, respectively. Across 10--200~Hz, signal gain decreases gradually while voltage-noise gain increases at both settings (Fig.~\ref{fig:afe}(b,c)). At 100~Hz, reducing $R_{\mathrm{in}}$ increases signal gain much more than voltage-noise gain. The 470-nF capacitor at the summing node contributes strongly to voltage-noise gain, and the separation between noise and signal gains is largest at $R_{\mathrm{in}}=330~\mathrm{k\Omega}$. The 330-k$\Omega$ feedback resistor and 2.2-nF capacitor give a nominal pole frequency of 219~Hz.

The calculated values form a partial budget of the op-amp stage, combining OPA392 voltage and current noise \cite{opa392ds} with resistor thermal noise in quadrature. For $R_{\mathrm{in}}=5.1$ and 330~\si{k\ohm}, the 10--200~Hz output-referred noise is 26.55 and 19.80~\si{\micro\volt} rms, and the input-referred noise is 0.511 and 24.88~\si{\micro\volt} rms. Op-amp voltage noise dominates, contributing 25.58 and 19.76~\si{\micro\volt} rms at the output. Higher gain lowers input-referred op-amp-stage noise but also amplifies signal and noise already at the microphone output. For the microphone, the ICS-40300 datasheet quotes a typical equivalent input noise of 31~dBA~SPL \cite{ics40300ds}, and Section~V evaluates the measured residual separately.

\section{Experimental Characterization}
\subsection{Bench protocol and recorded data}
\begin{figure}[!t]
\centering
\includegraphics[width=\columnwidth]{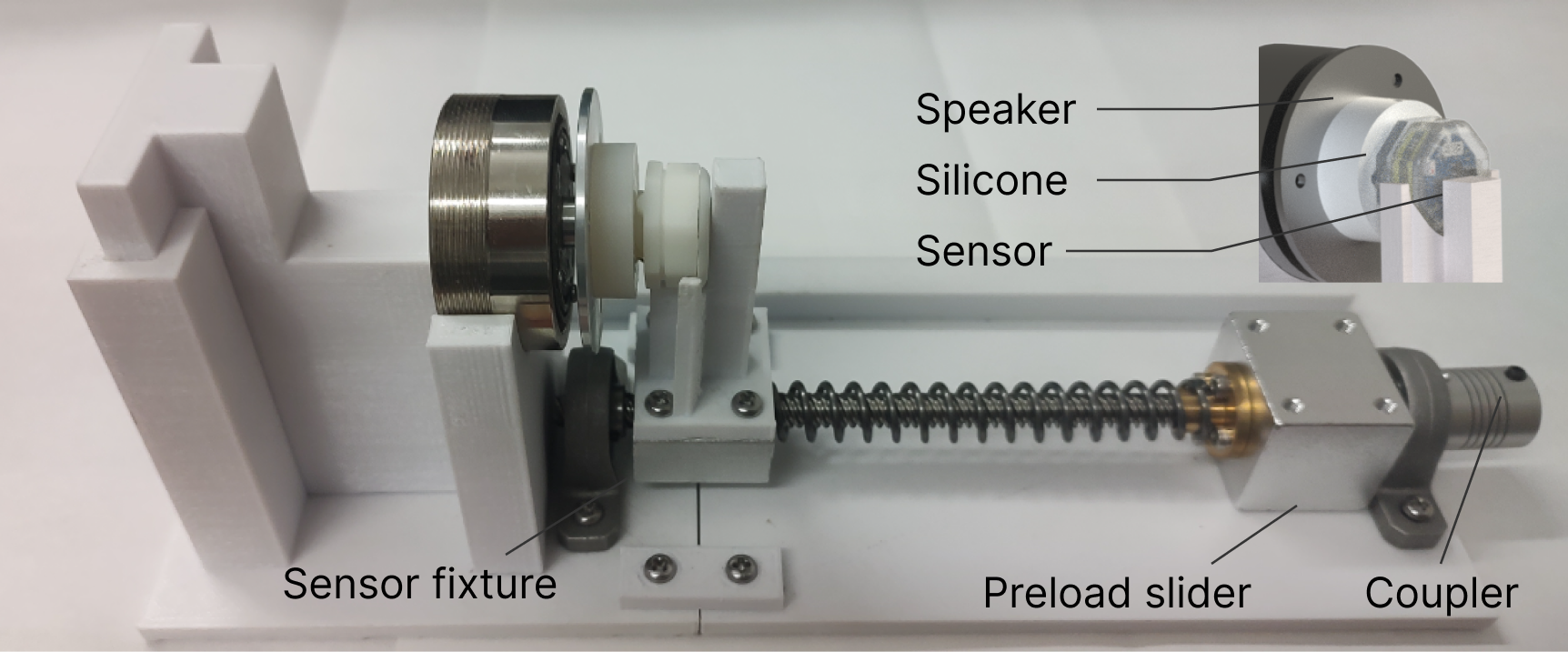}
\caption{Bench setup for frequency-sweep measurements. The photograph shows the sensor fixture, spring-loaded preload slider and manual adjustment coupler. The inset shows the speaker--silicone--sensor contact stack. Direct and Gap-ring recordings were paired within each housing at the same drive frequency, drive level and front-end gain.}
\label{fig:bench}
\end{figure}

In the fixture shown in Fig.~\ref{fig:bench}, a speaker mechanically excites a silicone layer (Shore A 20) placed on it. This layer forms the contact surface and stands in for skin and soft tissue. A manually adjusted spring-loaded slider sets the contact preload, and the fixture maintains alignment during each sweep. Each housing was tested under Direct and Gap-ring conditions at low and high drive levels, which are two speaker drive amplitudes specified as relative commands. The drive frequency was stepped from 10 to 1000~Hz in 10-Hz increments. At each frequency, we allowed a 2-s settling period before acquiring a 5-s recording. This protocol yielded 800 waveform audio (WAV) recordings spanning two housings, two contact conditions, two drive levels and 100 frequencies.

For the bench sweeps, the input resistance $R_{\mathrm{in}}$ (a trim potentiometer setting the front-end gain) was adjusted separately for each housing to obtain roughly similar recorded amplitudes. Each setting was then held fixed between the Direct and Gap-ring conditions, allowing contact dependence to be evaluated by paired attenuation within each housing. Absolute levels in decibels relative to full scale (dBFS) across housings also depend on the selected gains. The useful gain range is constrained by input-referred noise, reference noise and the ADC input range.

The recordings were single-channel, 40-kHz, 16-bit pulse-code modulation (PCM) streams. We divided the signed samples by 32768, removed the mean, applied a Hann window $w_n$ and scaled the single-sided fast Fourier transform (FFT) magnitude by $2/\sum_n w_n$. The drive-frequency FFT-bin magnitude, referenced to a digital full-scale peak of one, was the primary amplitude measure. Each 5-s record gave 0.2-Hz frequency spacing.

The analysis used 20 drive frequencies from 10 to 200~Hz, which span the low-frequency band where MMG content mainly lies \cite{orizio1993muscle,islam2013mechanomyogram}, yielding 160 recordings. Each combination of housing, contact condition, drive level and frequency had one recording. Medians describe the variation across the frequency sweep.

\subsection{Drive-frequency amplitude and contact attenuation}
\begin{figure}[!t]
\centering
\includegraphics[width=\columnwidth]{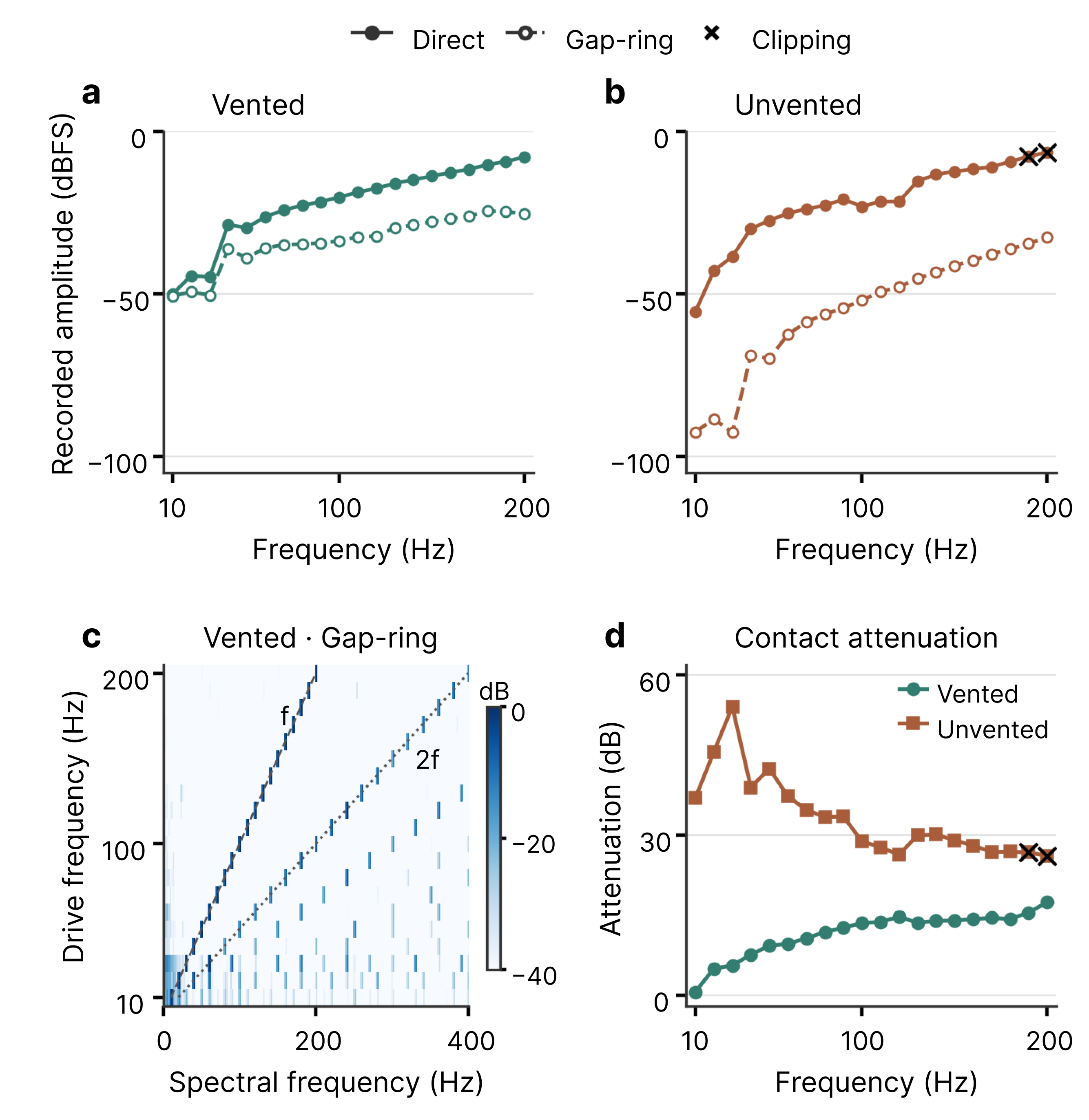}
\caption{Measured drive-frequency amplitudes and contact attenuation at the high drive level. (a,b) Drive-frequency amplitudes in dBFS for Direct and Gap-ring contact in the vented and unvented housings, respectively. (c) Vented Gap-ring spectra, each normalized to its maximum within 0--400~Hz. For display, the maximum spectral magnitude in each \mbox{2-Hz} interval is retained. The fundamental and second harmonic are marked. (d) Paired Direct-to-Gap-ring attenuation within each housing at fixed front-end gain. Crosses mark the two clipped unvented-Direct recordings at 190 and 200~Hz.}
\label{fig:response}
\end{figure}

For housing $h$, the within-housing contact attenuation is
\begin{equation}
A_h(f)=20\log_{10}\left|\frac{Y_{h,\mathrm{Direct}}(f)}{Y_{h,\mathrm{Gap}}(f)}\right|,
\end{equation}
where $Y$ is the recorded amplitude at the drive frequency. Fig.~\ref{fig:response} shows the high drive level, with panels (a) and (b) giving the recorded amplitudes in dBFS at each housing's front-end gain. Across 10--200~Hz, the median paired attenuation was 13.56~dB for the vented housing and 30.05~dB for the unvented housing, compared with 13.50 and 36.91~dB at the low drive level. The vented housing had lower attenuation at all 20 frequencies at both drive levels. At the high drive level, its attenuation increased from 9.3 to 14.3~dB over 50--180~Hz, whereas the unvented attenuation decreased from 42.5 to 26.9~dB.

At the low drive level, the ordering held over the nine frequencies where the Gap-ring fundamental exceeded the in-band residual in both housings (medians 14.74 and 34.28~dB). The ordering also held when the two clipped unvented-Direct recordings were excluded and when the largest peak within $\pm1$ or $\pm5$~Hz was selected.

\subsection{Fundamental-to-residual ratio}
To assess how clearly the imposed vibration remained detectable under each contact condition, we compared power at the drive fundamental with the remaining in-band power. We computed a Hann-window periodogram with single-sided power normalization and integrated spectral power using the 0.2-Hz bin spacing. Fundamental power was integrated within $\pm1$~Hz of the drive frequency. Residual power was integrated over 10--200~Hz after excluding $\pm1$~Hz around the fundamental and each integer harmonic within the band. The harmonics were excluded because the vented Gap-ring spectra in Fig.~\ref{fig:response}(c), where the drive fundamental dominates at most drive frequencies, also contain weaker components at its integer harmonics. The retained bandwidth ranged from 148.2 to 189~Hz across drive frequencies. The fundamental-to-residual ratio is the fundamental power divided by the residual power, expressed in decibels. A ratio of 0~dB denotes equal fundamental and residual power. The residual includes excitation-related components, interference and recording noise.

We computed the fundamental-to-residual ratio for all 160 recordings from 10 to 200~Hz. At the low drive level, the vented-housing medians were 20.56~dB under Direct contact and 8.59~dB with the Gap-ring, compared with 27.14 and $-2.54$~dB for the unvented housing. At the high drive level, the medians were 35.89 and 28.67~dB for the vented housing and 36.71 and 19.57~dB for the unvented housing. The Gap-ring thus caused a larger median reduction in the unvented housing at both drive levels. With the Gap-ring, the fundamental stayed above the residual at 18 of 20 low-drive and all 20 high-drive frequencies in the vented housing but fell below it at 11 low-drive and four high-drive frequencies in the unvented housing.

Excluding the two clipped unvented-Direct recordings and widening the exclusion bands from $\pm1$ to $\pm2$~Hz left the housing ordering unchanged, and all 11 below-zero unvented Gap-ring results at the low drive level were retained under the wider bands.

\Needspace{4\baselineskip}
\subsection{On-body waveform comparison}
We compared movement-aligned envelopes and spectra between housings during the same paced elbow movement, using the front-end settings in Section~IV. One healthy adult volunteer completed four trials per housing at one marked site over the brachioradialis. The housings were tested sequentially and refitted between trials. The analog output was sampled at 1~kHz with 16-bit resolution. Each housing retained its input resistance, keeping the signal within the ADC range. An audio-cued animation guided each trial through an initial rest period, ten elbow flexion--extension cycles and a final rest period. Each cycle comprised 1.5~s of flexion to approximately 90$^{\circ}$, a 3-s hold and 1.5~s of extension. Excluding the first and last cycles left eight per trial and 32 per housing. The rest level was computed over a 5-s segment from the rest periods.

We compared envelope timing and spectral shape using recordings normalized separately for each trial. Signals were band-pass filtered from 5 to 100~Hz using a zero-phase fourth-order Butterworth filter. Envelopes were calculated using a \mbox{50-ms} moving root-mean-square (RMS) window and normalized by the median peak across repetitions within each trial. We extracted windows from $-0.2$ to 2.0~s relative to the onset of each flexion and extension cue. Spectra were estimated using Welch periodograms of the 2-s post-cue windows.

\begin{figure}[!t]
\centering
\includegraphics[width=\columnwidth]{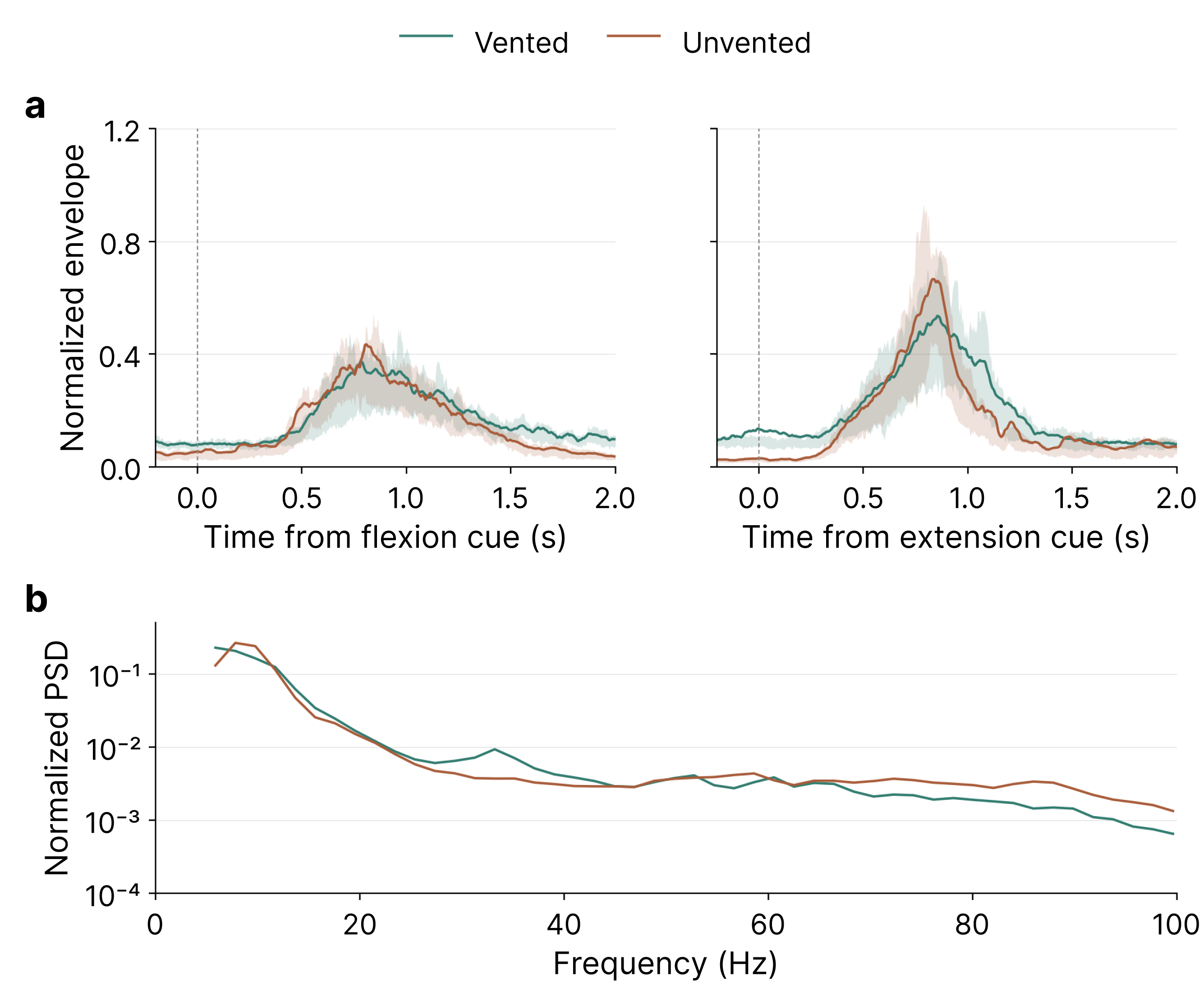}
\caption{On-body signal envelopes and spectra during paced elbow flexion and extension in one participant. (a) Mean envelopes (lines) and interquartile ranges across repetitions (shading), aligned to flexion cues on the left and extension cues on the right. Dashed vertical lines mark cue onset. (b) Power spectral density (PSD) over 5--100~Hz, shown on a logarithmic scale and normalized to unit integrated power within this band. Green and brown denote the vented and unvented housings, respectively.}
\label{fig:onbody}
\end{figure}

Both housings produced movement-aligned envelope bursts after the flexion and extension cues (Fig.~\ref{fig:onbody}). Flexion peaks occurred 0.78 and 0.81~s after the cue for the vented and unvented housings, respectively, and extension peaks at 0.85 and 0.83~s. The housing-mean envelopes and log spectra correlated at $r=0.935$ and $r=0.956$, respectively. Across trials, mean correlations within the same housing were 0.891 for envelopes and 0.985 for log spectra, compared with 0.878 and 0.948 across housings. The across-housing envelope correlation was close to the within-housing value, so the housing difference was comparable to the variation from refitting.

During the 5-s rest segments, the 5--100~Hz RMS level was 4--9 ADC counts, with signal levels in the 2-s post-cue windows 8--19~dB above rest. Both housings had a median frequency of 9.8~Hz. The fraction of power in the 40--100~Hz band was 7.3\% for the vented housing and 9.5\% for the unvented housing. The two housings showed similar envelope timing and broadly similar spectra for this participant.

\section{Discussion}
The side vent adds a parallel leakage path that carries most flow, so cavity pressure is less sensitive to rim clearance. Without the vent, an opening rim becomes the main leakage path and sharply lowers microphone pressure. Larger vents give diminishing returns, as each doubling of area narrowed the open-clearance span similarly at a growing gain cost. Vent area sets this gain cost much as vent size sets low-frequency reduction in earmolds \cite{stuart1992probe,kuk2009comparison}. Earplug leaks limit attenuation \cite{henriksen2008using} and vents relieve occlusion \cite{denk2023occlusion}. This vent instead reduces dependence on rim fit.

Geometric comparisons show that ideal-closure gain alone does not characterize performance under clearance. Eleven modest variants changed ideal-closure gain by up to about 3~dB but open-clearance gain by at most 0.7~dB. Reducing cavity height raises gain more at ideal closure than at \mbox{0.20-mm} clearance. Lateral scaling can raise gain there yet increase the open-clearance span. Earlier chamber studies selected geometry for gain and frequency-response flatness \cite{watakabe2001mmg,posatskiy2012chamber}. Because attachment stability matters for wearable MMG \cite{jin2026multisensor}, housing design should weigh contact dependence alongside acoustic gain and frequency-response flatness.

The simulations showed lower sensitivity to rim clearance for the vented cavity, while the paired bench measurements showed lower Direct-to-Gap-ring attenuation for the assembled vented sensor. The simulation prescribes source motion, whereas the Gap-ring in the bench test can change both acoustic leakage and the motion delivered to the cavity. In the linear regime, the recorded amplitude is $Y_h=G_hH_hu_h$, with gain $G_h$, acoustic transfer $H_h$ ($H$ in (1)), and source displacement $u_h$ at the cavity. With the front-end gain $G_h$ held fixed within each housing, contact attenuation becomes
\begin{equation}\begin{split}
A_h={}&20\log_{10}\left|\frac{H_{h,\mathrm{Direct}}}{H_{h,\mathrm{Gap}}}\right|\\
&+20\log_{10}\left|\frac{u_{h,\mathrm{Direct}}}{u_{h,\mathrm{Gap}}}\right|.
\end{split}\end{equation}
The first term represents the change in acoustic gain and the second the change in mechanical input, consistent with the reported influence of normal force on MMG acquisition \cite{castillo2021mmgforce}. If both housings, which share the Gap-ring and fixture, show a similar change in mechanical input, their attenuation difference arises mainly from the simulated acoustic term. The simulations identify the acoustic mechanism, and the paired measurements quantify the assembled sensor's response.

With the Gap-ring at the low drive level, fundamental power fell below residual power more often for the unvented housing. The Gap-ring lowered the median fundamental-to-residual ratio by about 12~dB in the vented and 30~dB in the unvented housing. These drops follow the paired attenuations, so detectability under degraded contact depends mainly on the resulting signal loss. The vented housing kept the fundamental detectable despite its lower acoustic gain. Front-end gain restores the ADC signal level at a small calculated op-amp noise cost (Section~IV-B), and acoustic gain sets the signal-to-noise ratio relative to microphone self-noise.

\section{Conclusion}
This study examined side venting to reduce contact dependence in microphone-based MMG housings. Finite-element simulations at 100~Hz gave an open-clearance span of 10.0~dB with venting and 25.8~dB without, which branch-flow analysis attributed to a parallel pressure-relief path. At the high bench drive level, median paired attenuation was 13.6~dB for the vented and 30.1~dB for the unvented housing. With the Gap-ring, the vented housing kept the fundamental above the residual at 18 of 20 low-drive and all 20 high-drive frequencies, whereas the unvented fundamental fell below it at 11 low-drive frequencies. A single-participant recording during paced elbow flexion and extension showed similar envelope timing and broadly similar spectra for both housings.

Contact dependence should therefore be a housing design criterion alongside acoustic gain and frequency-response flatness. Front-end gain restores the ADC signal level lost to venting at a small calculated op-amp noise cost, and acoustic gain sets the signal-to-noise ratio relative to microphone self-noise. Each doubling of vent area lowered contact dependence by similar amounts at a growing gain cost, giving diminishing returns.

Current evidence comprises single-participant on-body data, simulations with prescribed source motion, and an op-amp noise calculation. Future work will extend on-body evaluation to multiple participants with repeated donning and voluntary movement, measure ambient sound and airflow picked up through the vent, and test the separation of acoustic and mechanical terms in~(7) with bench-measured contact-surface displacement. Full readout noise, including the microphone, will be measured end to end to set vent area against available readout gain. Vented housings will be integrated into exoskeleton control and electrical-stimulation feedback systems to assess performance under continued use.

\bibliographystyle{IEEEtran}
\bibliography{refs}
\end{document}